\documentclass[onecolumn,showpacs,preprintnumbers,amsmath,amssymb]{revtex4}

\usepackage{graphicx}
\usepackage{dcolumn}
\usepackage{bm}
\usepackage{subfigure}

\begin{document}


\title{Do consistent $F(R)$ models mimic General Relativity plus
$\Lambda$?}

\author{Stephen A. Appleby and Richard A. Battye \\ {\it Jodrell Bank
Observatory, Department of Physics and Astronomy,} \\ {\it
University of Manchester, Macclesfield, Cheshire, SK11 9DL}}


\date{\today}

\begin{abstract}

Modified gravity models are subject to a number of
consistency requirements which restrict the form that the function $F(R)$ 
can
take. We study a particular class of $F(R)$ functions which satisfy
various constraints that have been found in the literature. These
models have a late time accelerating epoch, and an acceptable matter
era. We calculate the Friedmann equation for our models, and show
that in order to satisfy the constraints we impose, they must
mimic General Relativity plus $\Lambda$ throughout the cosmic history, with
exponentially suppressed corrections. We also find that the free
parameters in our model must be fine tuned to obtain an acceptable
late time accelerating phase. We discuss the generality of this conclusion.

\end{abstract}

\maketitle

\section{Introduction}

In recent years, experimental cosmology has provided strong evidence
that the Universe is currently undergoing a phase of accelerated
expansion. The physical phenomena responsible for this expansion has
proved elusive, and there are numerous theories that attempt to
explain the current epoch of the Universe. Typically, the observed
acceleration is attributed to a new energy component, dark energy,
which dominates at late times. Numerous models
exist of this form, the simplest of which include introducing a
cosmological constant $\Lambda$, or alternatively a scalar field slowly
rolling down a potential.

Although dark energy is undoubtedly the most popular explanation of
the current epoch of the Universe, it is not the only way to obtain
late time acceleration. In this paper we consider a class of models
where gravity is modified at large scales, in such a way that the
late-time expansion of the Universe could arise naturally from the
new gravitational field equations. These modified gravity models
have been considered by numerous authors, see for example
\cite{sk1,wd1,sl1,d1,cr4,db1,so1,teg1,cr6,at10,fa1,az1,cap1,no10,no11,dv6,ee1,sl2,c1,am1,ca1,cr3,df1,n1,f1,f2,zh1}.

Modifying gravity in a manner consistent with experimental data has
proved difficult. General Relativity (GR) is a very robust and well
tested theory, and it has been found that even slight modifications
often lead to instabilities, such as the propagation of ghosts
\cite{sl1}, or in otherwise stable matter sources
\cite{d1},\cite{wd1}. When introducing additional terms into the
gravitational action, we must be careful to respect the success of
GR in both the low and high curvature regimes, to
ensure that any potentially new model agrees with known
observational tests of gravity.

In the literature, there are numerous simple conditions that exist
which restrict the form that a modified gravity model may take. In
this paper, we combine all of these conditions, and in doing so find
that the modified gravity function $F(R)$ is well constrained, even
before considering stringent experimental tests. The philosophy
adopted throughout is that before attempting to reconcile any
potential modified gravity model with experimental data, that
modified gravity model must first satisfy these consistency requirements.

We will take a trial $F(R)$ model which satisfies all of the
existing constraints. We find that our model exhibits a late-time 
accelerating solution, in the absence of a
cosmological constant. We then consider the cosmology of our model,
and obtain a modified Friedmann equation. We show that to satisfy all
constraints, and to obtain an acceptable Newtonian limit for small
$R$, our model must act like GR with highly
suppressed corrections throughout its cosmological history. In
addition, we find that although our model has no true cosmological
constant, we must still fine tune a particular combination of the
free parameters in $F(R)$ to obtain an acceptably small curvature in
the current accelerating epoch. We conjecture that our results are generic 
features
of models which satisfy the consistency requirements.

\section{\label{sec:1} Relevant Field equations in the Jordan and Einstein
frames}

\noindent Before we consider a specific $F(R)$ model, we first
briefly review the field equations in modified gravity theories.
Throughout this paper, we will study the action

\begin{equation}\label{eq:ab1} S =  {M^{2} \over 2}\int \sqrt{-g} d^{4}x
F(R) + S_{{\rm m}},\end{equation}

\noindent where $F(R)$ is some function of the Ricci scalar $R$, $M$
is the mass scale of our model, and $S_{{\rm m}}$ the matter action.
The field equations obtained by varying the action ($\ref{eq:ab1}$)
with respect to the metric are

\begin{equation}\label{eq:b2} F'(R) R_{\mu\nu} - {1 \over 2} F(R)g_{\mu\nu}
-
\nabla_{\mu}\nabla_{\nu}F'(R) + g_{\mu\nu} \Box F'(R) = {1 \over
M^{2}}T_{\mu\nu} ,\end{equation}

\noindent where $T_{\mu\nu}$ is the energy momentum tensor of any
matter present. Taking the trace of ($\ref{eq:b2}$) gives the more
useful equation

\begin{equation}\label{eq:b3} Q(R) + 3 \Box F'(R) = {T \over M^{2}}
, \end{equation}

\noindent where the function $Q(R)$ is given by $Q(R) = RF'(R) -
2F(R)$. Vacuum solutions to the field equations satisfy $R = R_{0} =
{\rm const}$ when $T = 0$, and hence
$Q(R_{0})=0$.

In addition to the above field equations in the Jordan frame, we
will also want to consider the potential of the scalar field in the
Einstein frame. To transform to the Einstein frame, we first write
($\ref{eq:ab1}$) as

\begin{equation} \label{eq:2} S = {M^{2} \over 2}\int \sqrt{-g} d^{4}x\left[
F'(\phi) \left( R
- \phi \right) + F(\phi) \right]+ \int \sqrt{-g} d^{4}x{\cal L}_{\rm m} ,
\end{equation}

\noindent where ${\cal L}_{\rm m}$ is the matter Lagrange density. The
actions ($\ref{eq:2}$) and ($\ref{eq:ab1}$) are equivalent if
$F''(R) \neq 0$. We then use the conformal transform
$\tilde{g}_{\mu\nu} = F'(\phi) g_{\mu\nu}$, which gives

\begin{equation} R = F'(\phi) \left( \tilde{R} + 3\tilde{\Box}\ln F' -
{3 \over 2 (F')^{2}}\tilde{\nabla}_{\alpha} F'
\tilde{\nabla}^{\alpha} F'  \right) ,\end{equation}

\noindent where $R$ and $\tilde{R}$ are the Ricci scalars associated
with the metrics $g_{\mu\nu}$ and $\tilde{g}_{\mu\nu}$ respectively.
Using this in ($\ref{eq:2}$) gives the following Einstein frame
action

\begin{equation} S = \int \sqrt{-\tilde{g}} d^{4}x \left[
{M^{2} \over 2}\tilde{R}
- {3M^{2} \over 4 (F')^{2}}\tilde{\nabla}_{\alpha} F'
\tilde{\nabla}^{\alpha} F' - V(F'(\phi)) \right] + \int \sqrt{-\tilde{g}} 
d^{4}x  {1 \over (F')^{2}}{\cal
L}_{\rm m} ,\end{equation}

\noindent where the Einstein frame potential is given by

\begin{equation} V(\phi) = {M^{2} \over 2}\left({\phi F'(\phi) - F(\phi) 
\over
F'(\phi)^{2}}\right) .\end{equation}

Finally, by making the field redefinition $F'(\phi) = \exp(\sigma
\sqrt{2} / \sqrt{3M^{2}})$, we obtain

\begin{equation}\label{eq:4} S = \int \sqrt{-\tilde{g}} d^{4}x \left[
{M^{2} \over 2}\tilde{R}
- {1 \over 2}\left(\tilde{\nabla}\sigma\right)^{2} - V(\sigma) + {1
\over (F')^{2}}{\cal L}_{\rm m}\right],
\end{equation}

\noindent with potential

\begin{equation}\label{eq:ab5} V(\sigma) = {M^{2} \over 2}\left[\phi(\sigma) 
e^{\sigma
\sqrt{2} /
\sqrt{3M^{2}}} - F(\phi(\sigma))\right]e^{-2 \sigma\sqrt{2} /
\sqrt{3M^{2}}}.
\end{equation}

\noindent In many instances, it is preferable to study modified
gravity in the Einstein frame, since the field equations are much
simpler. Additionally, the potential ($\ref{eq:ab5}$) is useful
since its critical points correspond to the vacuum solutions $Q(R_{0}) =
0$ in the Jordan frame. We stress that the two frames are
globally equivalent only if $F''(R) \neq 0$.

\section{\label{sec:2}Review of constraints on $F(R)$ models}

Here we briefly review the constraints in the literature imposed on
the form of the function $F(R)$. We begin by demanding $F'(R) > 0$
and $F'' \neq 0$ for all $R$. The first of these is to ensure stability; it 
has
been shown \cite{sl2} that if $F'(R)<0$ for any $R$ then the model
will possess ghost instabilities in this region. The second
condition $F''(R) \neq 0$ is imposed since it has been found that models 
which have $F''(R)<0$ will generically possess instabilities,
when we consider the metric variational approach of our model \cite{so1}. We 
note that the possibility of $F''(R)$ being zero for some $R$
has not been completely discounted in the literature. Nevertheless, we will
impose $F'' \neq 0$ for all $R$.

The next constraint that we impose on $F(R)$ is that in the limit $R
\to \infty$, we have $F(R) / R \to 1$. Writing $F(R) = R +
\epsilon(R)$, this implies that $\epsilon / R \to 0$ as $R \to
\infty$. This in turn implies that $F'(R) \to 1$ as $R \to \infty$,
from which we can deduce that $\epsilon(R) \to {\rm const}$ as $R \to 
\infty$.
This condition is imposed to ensure that the modified gravity model
acts suitably like $\Lambda {\rm CDM}$ in the large $R$ regime,
where gravity is well constrained by PPN constraints and the CMB.

With this bound, we have $F'(R) \to 1$ as $R \to \infty$. However,
in ref. \cite{hu1} it was found that for a given $F(R)$ theory to be
stable to perturbations in the large $R$ regime, we must have
$F''>0$ as $R \to \infty$. This means that $F'(R)$ must asymptote to
unity from below, since $F''(R)>0$ implies that it must be an
increasing function. Next, we can use the fact that we have imposed
$F''(R) \neq 0$ for all $R$. This means that there can be no turning
points in $F'(R)$, implying that if $F'(R)$ asymptotes to unity from
below, then it must remain less than unity for all $R$. However, we
have also imposed $F'(R)>0$. From this we conclude that $F'(R)$ must
be an increasing function of $R$, and must satisfy $0 < F'(R) < 1$.

Another set of constraints on $F(R)$ are discussed in ref.
\cite{am1}, where the cosmology of $F(R)$ models was considered. In
order for a theory to be a viable alternative to GR,
it must exhibit an acceptable cosmological history, that is it must
have suitable radiation, matter and late time accelerating eras. It
was found that in order for a particular $F(R)$ to have a valid
matter era, $F(R)$ must satisfy $m(r) \approx +0$ and ${dm \over dr}
>-1$ at $r=-1$, where $m(r)$ and $r$ are given by $m = R F'' / F'$ and $r
= - R F' / F$. In addition, to obtain a late time accelerating era such as 
we are
currently in, the function $m$ must satisfy either; $(a)$ $m=-r-1$,
$(\sqrt{3}-1)/2 < m \leq 1$ and $dm/dr < -1$ or $(b)$ $0 \leq m \leq
1$ at $r=-2$. If these are not satisfied for any $R$,
then the model will not have an acceptable cosmological history.

\section{\label{sec:3} A trial $F(R)$ model}

In this section, we give an example $F(R)$ function which satisfies
all of the conditions in the previous section. Given that $F'(R)$
must be an increasing function, and lie in the range $0<F'(R)<1$, an
obvious choice is

\begin{equation}\label{eq:fg1} F'(R) = {1 \over 2} \left[ 1 + \tanh(aR-b) 
\right],
\end{equation}

\noindent where $a>0$ and $b$ are free parameters. This function has
the appropriate behaviour $F'(R) \to 1$ as $R \to \infty$ and $F'
\to 0$ as $R \to -\infty$. Integrating this $F'(R)$ gives us

\begin{equation}\label{eq:ab2} F(R) = {R \over 2} + {1 \over 2a} \log
\left[\cosh(aR-b)\right] +
A ,\end{equation}

\noindent where $A$ is an integration constant. We
are looking for $F(R)$ models that exhibit late time
accelerating solutions in the absence of a cosmological constant. Therefore 
to specify $A$, we impose that $F(0) = 0$.
Hence we find $A = -(1 / 2a) \ln \cosh(c)$ and

\begin{equation}\label{eq:ab3} F(R) = {1 \over 2}R + {1 \over 2a} \log
\left[\cosh(aR) - \tanh(b)\sinh(aR) \right].\end{equation}

\noindent It is this $F(R)$ that will be studied for the remainder
of this paper.

\begin{figure}[htp]
     \centering
     \subfigure[]{
          \label{fig:A1}
          \includegraphics[scale=0.5]{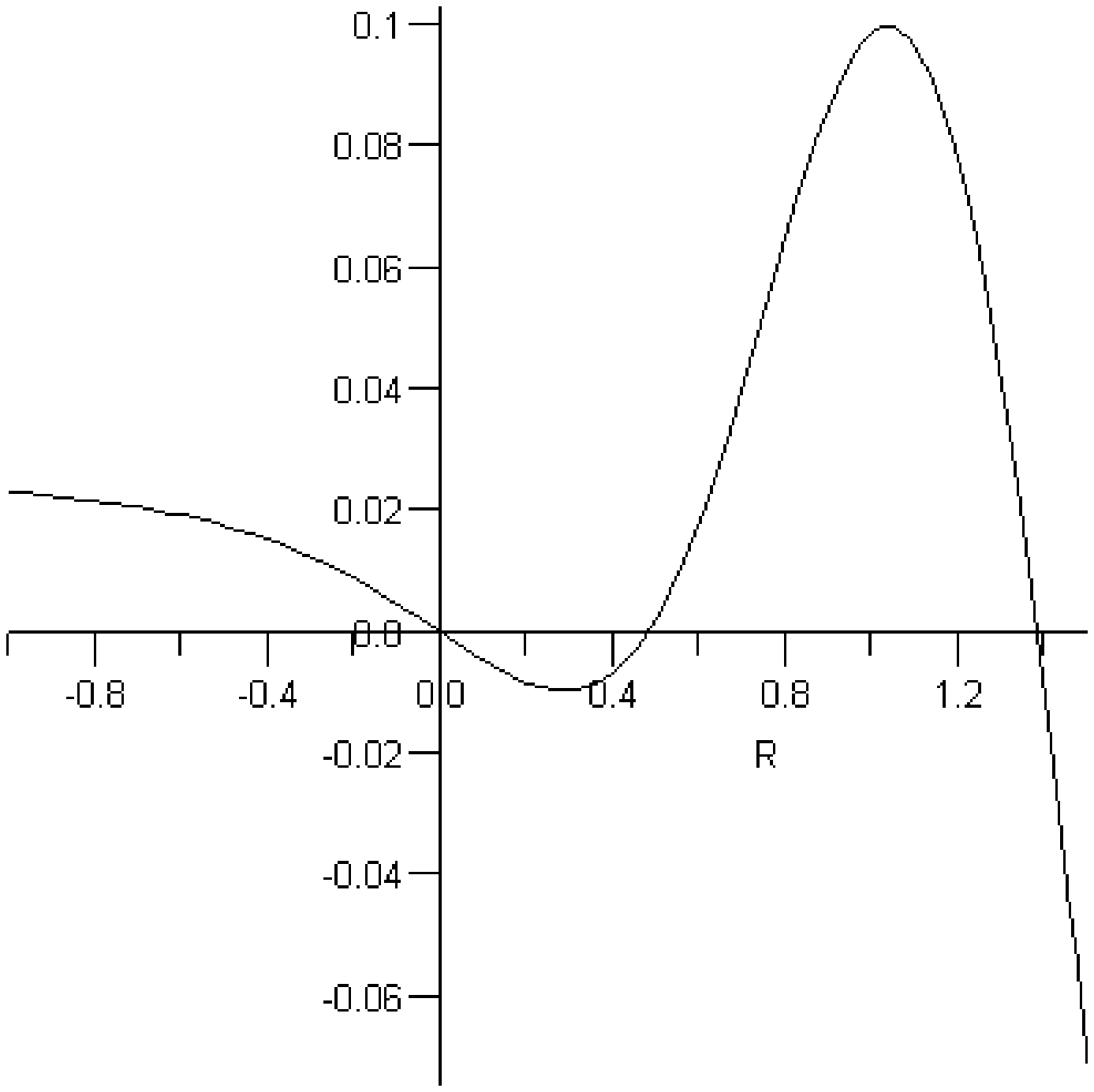}}
     \hspace{.2in}
     \subfigure[]{
          \label{fig:m}
                \includegraphics[scale=0.5]{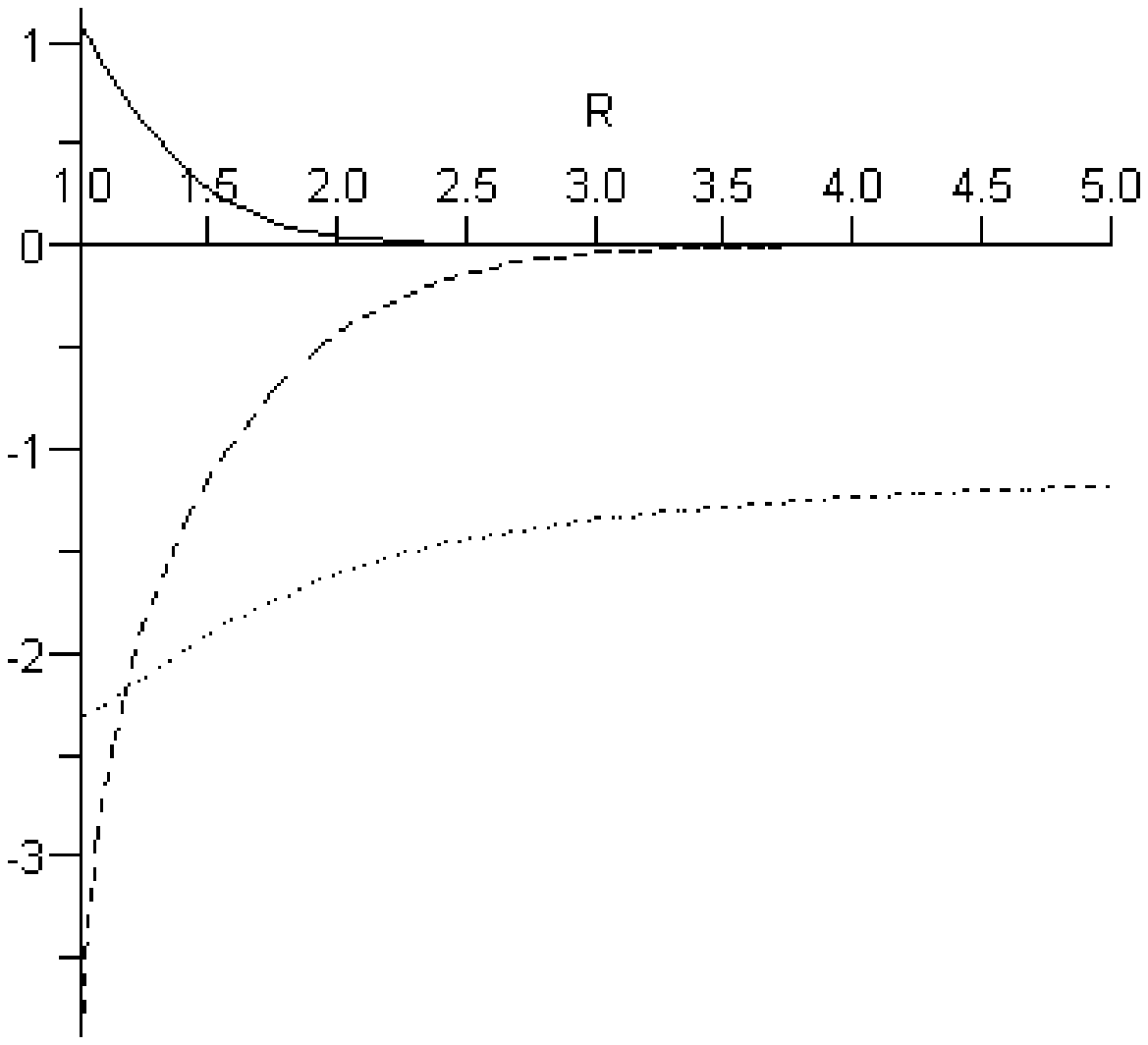}}
     \caption{$(a)$ is a plot of the function Q(R). This function has three
zeros, which correspond to three vacuum states of our model.
     $(b)$ are the functions $m$ (solid), $r$ (dotted) and $dm/dr$ (dashed),
plotted parametrically as functions of $R$. We can clearly see that there
exists an $R$
     such that $r=-2$ and $m<1$, suggesting a late-time accelerating phase.
In addition, we have $m \sim 0$ as $r \to -1$, which is the condition
required for an acceptable matter era \cite{am1}. We have set $a=2$,
     $b=1.5$.}
     \label{fig:2858multifig}
\end{figure}

We will begin by considering some of the general properties of our
model. First, we look for vacuum solutions to the field equations,
that is solutions to $Q(R_{0})=0$. The behaviour of $Q(R)$ is
exhibited in Fig. $\ref{fig:A1}$. We have found that $Q(R)$ for
our $F(R)$ generically possesses three zeros, for any $a$ and $b
\gtrsim 1.2$. This means that our $F(R)$ model can have three vacuum
states, which are Minkowski space ($R=0$), and two de-Sitter vacua, one of 
which is stable.

\begin{figure}[htp]
     \centering
     \subfigure[]{
          \label{fig:Vt}
          \includegraphics[scale=0.5]{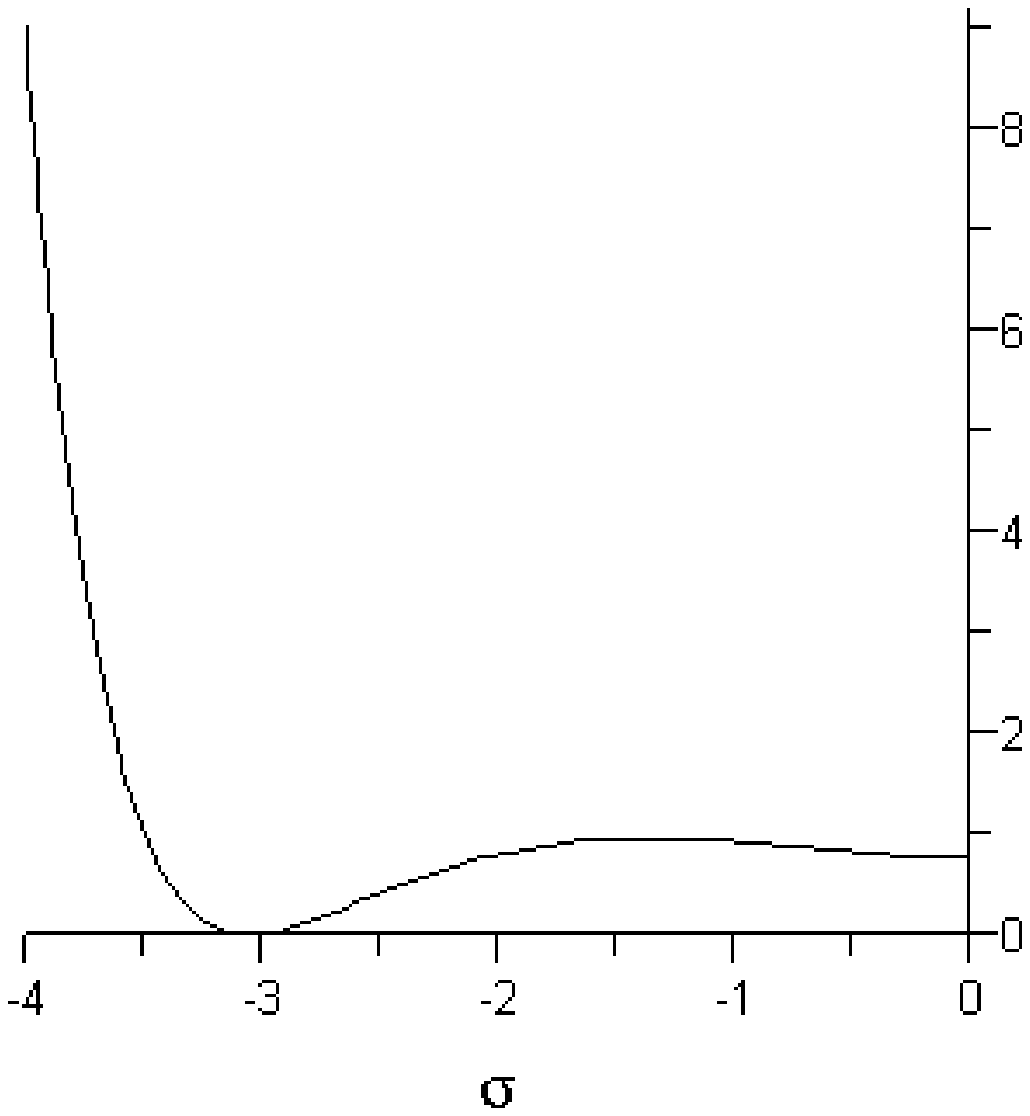}}
     \hspace{.2in}
     \subfigure[]{
          \label{fig:Vsmall}
                \includegraphics[scale=0.5]{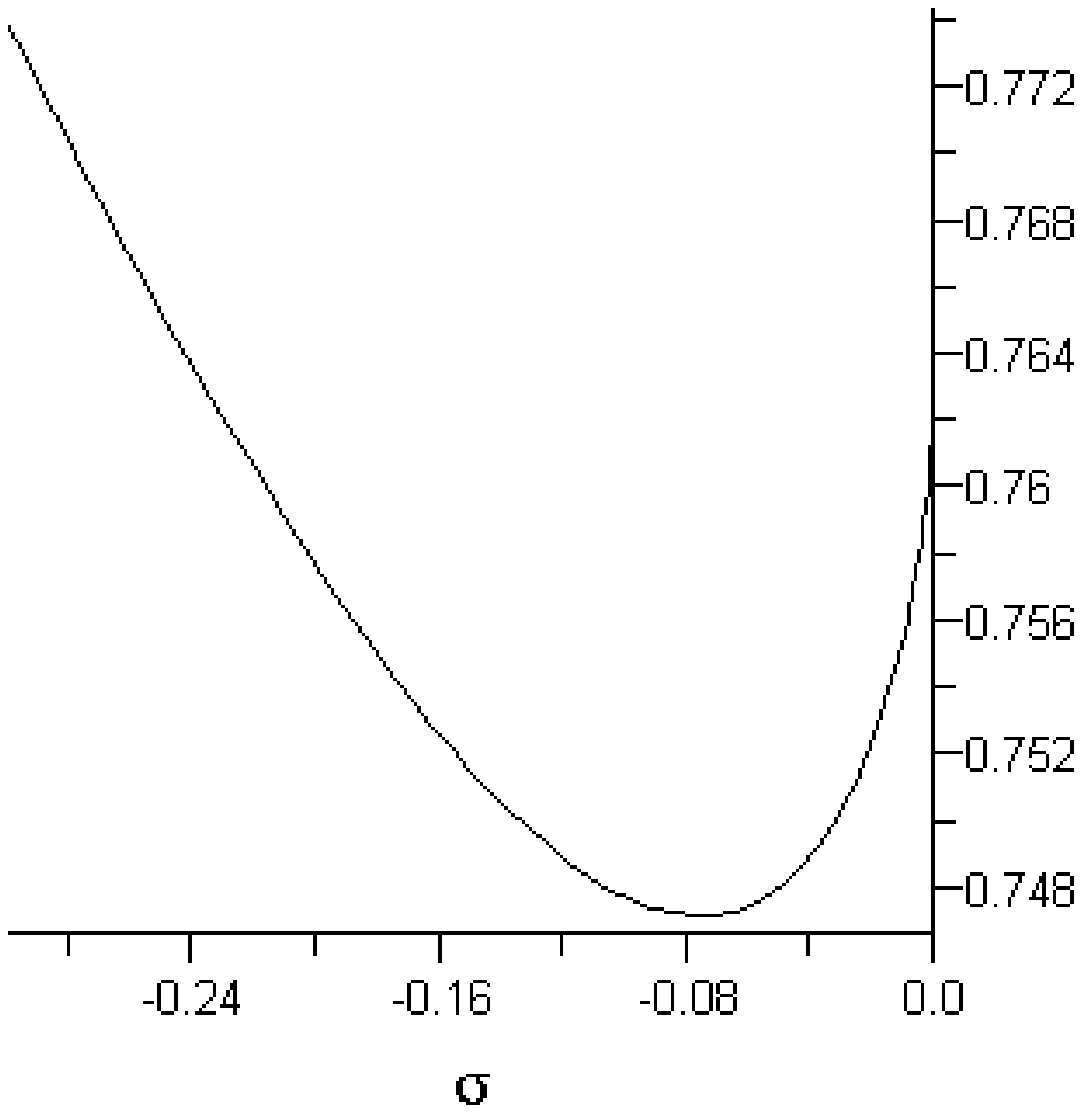}}
     \caption{$(a)$ The Einstein frame potential. Note the presence of two 
minima; one at $V=0$, and one at some $V = V_{0} > 0$. It is this minima
(shown in $(b)$) that corresponds to a stable de Sitter solution in the
Jordan frame. $(b)$ is $V(\sigma)$ in the
small $\sigma$ regime. We have set $a=2$, $b=1.5$.}
     \label{fig:2859multifig}
\end{figure}

To see these vacuum states more explicitly, we move to the Einstein
frame. A plot of the potential $V(\sigma)$ for our $F(R)$ is given
in Figs. $\ref{fig:Vt}$, $\ref{fig:Vsmall}$. We can clearly see that the 
potential has two
minima, one at $V = 0$ (which corresponds to the Minkowski vacuum)
and the other at some $V = V_{0}>0$. This second minima corresponds
to a stable de Sitter vacuum solution in the Jordan frame, and hence
our model has a stable accelerating solution, in the absence of a
cosmological constant. This is exactly the behaviour we seek. Finally, the 
third vacuum in the Jordan frame corresponds
to the maxima in the potential $\ref{fig:Vt}$, and is unstable.

Following \cite{am1}, we also briefly consider the functions $m(r)$,
$r$ and $dm/dr$ for our model, as given in section $\ref{sec:2}$. We
plot these functions parametrically as functions of $R$ in
Fig. $\ref{fig:m}$. We see that there is a region in which $m \approx
0$ and $dm/dr >-1$ for $r \approx -1$, which corresponds to a matter
era. As $R \to \infty$, we have $r \to -1$ and $m \to
0$, suggesting that the condition for a matter epoch is satisfied
asymptotically. Further, as $R$ decreases, there is a
region where $r = -2$, at which $0<m<1$. This region corresponds to
the late-time de-Sitter phase. Hence our model connects the points
$(-1,0)$ and $(-2, a<1)$ in the $(r,m)$ plane, as is required
\cite{at1} for it to have an acceptable cosmic history.

\section{\label{sec:4} Newtonian Limit}

We now consider the Newtonian limit of our model. In doing so, we
will be able to constrain the free parameters $a$ and $b$. The
constant $a$ will be fixed by the following two conditions; we
require the de-Sitter vacuum solution to be $R = R_{0} \approx
12H_{0}^{2}$, and as we will see shortly, we require $F''(R_{0}) \ll
F'(R_{0})$. Using these conditions, the constant $b$ will not be
fixed, but we will find that it must be large.

To see how $a$ and $b$ are determined, we consider the Newtonian
limit of modified gravity theories. The full field equations for the
action ($\ref{eq:ab1}$) are

\begin{equation}\label{eq:bm2} F'(R) R_{\mu\nu} - {1 \over 2} F(R)g_{\mu\nu}
-
\nabla_{\mu}\nabla_{\nu}F'(R) + g_{\mu\nu} \Box F'(R) = {1 \over
M^{2}}T_{\mu\nu} .\end{equation}

\noindent We would like to perform a weak-field expansion of these
equations around the de-Sitter minima of our model. In doing so, we
hope to obtain an acceptable Newtonian limit. An analysis has been
performed previously in Ref. \cite{di1}, and expanding the field
equations about a symmetric de-Sitter vacuum gives

\begin{equation} F'(R_{0})\delta R_{\mu\nu} - {1 \over
4}F''(R_{0})R_{0}g_{\mu\nu}\delta R - {1 \over 2}F(R_{0})\left(
\delta g_{\mu\nu} - {1 \over 2}g_{\mu\nu}\delta g\right) -
F''(R_{0}) \left(\nabla_{\mu}\nabla_{\nu}\delta R + {1 \over 2}
g_{\mu\nu} \Box \delta R \right) = {1 \over M^{2}} \left(T_{\mu\nu}
- {1 \over 2}g_{\mu\nu} T \right),
\end{equation}

\noindent where

\begin{equation} \delta R_{\mu\nu} = {1 \over 2} \left(
\nabla_{\mu}\nabla^{\sigma} \delta g_{\sigma\nu} +
\nabla_{\nu}\nabla^{\sigma}\delta g_{\sigma\mu}\right) + {1 \over
3}R_{0} \delta g_{\mu\nu} - {1 \over 12}R_{0}g_{\mu\nu}\delta g - {1
\over 2} \Box \delta g_{\mu\nu} - {1 \over
2}\nabla_{\mu}\nabla_{\nu}\delta g .\end{equation}

\noindent In order to obtain an acceptable Newtonian limit from
these perturbative equations, we require that $F'(R_{0}) \sim O(1)$,
$F(R_{0}) \sim O(R_{0}) \ll 1$ and $F''(R_{0}) \ll F'(R_{0})x^{2}$,
where $x$ is a macroscopic distance. If this is the case, then the
perturbative equations ($\ref{eq:b2}$) will be approximately given
by

\begin{equation}\label{eq:b3} F'(R_{0}) \delta R_{\mu\nu} \approx {1 \over
M^{2}}\left(T_{\mu\nu} - {1 \over 2}g_{\mu\nu}T \right).
\end{equation}

\noindent Equation ($\ref{eq:b3}$) is simply the weak field
expansion of General Relativity, from which the standard Newtonian
limit is recovered. The only difference is the coefficient of the
first term in ($\ref{eq:b3}$), which is $F'(R_{0}) \sim O(1)$ as
opposed to unity. Note that to obtain the correct limit we must have
$F''(R_{0}) \ll F'(R_{0})$; if this were not the case then fourth order
derivative terms such as $\nabla_{\mu}\nabla_{\nu}\delta R$ would be
present in ($\ref{eq:b3}$), and it has been shown that these terms
would lead to strong curvature at all lengthscales \cite{sx1}.

For our model (and generically in $F(R)$ theories), we have
$F(R_{0}) \sim O(R_{0}) \ll F'(R_{0})$ for $R_{0} \sim H_{0}^{2}$,
which is the first condition to obtain an acceptable Newtonian
limit. However, the second condition, $F''(R_{0}) \ll F'(R_{0}) x^{2}$,
is not generically satisfied, and we must choose our free parameters
to obtain the correct behaviour. From ($\ref{eq:fg1}$), we find
$F''(R)$ to be

\begin{equation}\label{eq:ac1} F''(R) = {a \over 2} {\rm sech}^{2}(aR -
b)  ,\end{equation}

\noindent and hence to ensure that $F''(R_{0})$ is very small, we
require $aR_{0} - b \gg 1$. We can use this condition to fix $a$ in
terms of $R_{0}$ and $b$. To do so, we note that in the vicinity of
the de-Sitter minima, $R \sim R_{0}$, we can expand $F(R)$ and
$F'(R)$ for large $\alpha = aR-b$,

\begin{align}\label{eq:fb3} F(R) &\approx R - {\log \left(1+e^{2b}\right) \over 2a} + {1 \over 2a} 
e^{-2\alpha}  + O\left(e^{-4\alpha}\right) \\ F'(R) &\approx 1 -
e^{-2\alpha} + O\left(e^{-4\alpha}\right) \\
F''(R) &\approx 2 a e^{-2\alpha} + O\left(e^{-4\alpha}\right),
\end{align}

\noindent and hence from the vacuum field equations $Q(R_{0})=0$, we
can use our approximations ($\ref{eq:fb3}$) to give

\begin{equation} R_{0} = 12 H_{0}^{2} \approx {1 \over a}
\log\left(1+e^{2b}\right) ,\end{equation}

\noindent from which we find $a \approx {1 \over
R_{0}}\log\left(1+e^{2b}\right)$. This fixes the free parameter $a$
in terms of $R_{0}$ and the remaining free parameter $b$. Next, we
can constrain $b$ from the condition $F''(R_{0}) \ll F'(R_{0}) \approx 1$. Using
this, and expanding $F''(R)$ in powers of
$e^{-2\alpha}$, we find that $b$ must satisfy

\begin{equation} {\log (1+e^{2b}) \over \cosh^{2}b} \ll 2R_{0},
\end{equation}

\noindent which is satisfied for large $b$, such that $8be^{-2b} \ll
2R_{0}$.

To review our conditions; we are working with a model with two free
parameters, $a$ and $b$, and we have imposed two constraints on this
model, which are that $R_{0} = 12H_{0}^{2} \ll 1$ and $F''(R_{0})\ll 1$.
These constraints force us to choose $a$ and $b$ such that $a
\approx {1 \over 12 H_{0}^{2}}\log\left(1+e^{2b}\right)$ and
$8be^{-2b} \ll 2R_{0}$. Together, they imply that $b \gg 1$, and hence
$a \approx {2 b \over R_{0}}$. We note that we have not been forced
to specify the constant $b$, other than imposing the condition
$b \gg 1$. This parameter controls how closely our model mimics GR; the 
larger we set $b$ the more suppressed our
corrections $\exp(-2\alpha)$ become.

Before moving onto the cosmology of our model, we remark that the
above weak-field expansion was performed around the de-Sitter vacuum
$R = R_{0}$. It has been noted  \cite{zh1} that for
local tests of gravity, the expansion should instead be around the
local energy density of the solar system, which is much larger than
$R_{0}$. In a subsequent paper we will perform a detailed analysis
of the local tests of gravity for our model, but for now we simply
expand around $R=R_{0}$. Performing a weak-field expansion around a
solar system background will not significantly modify our conclusion;
we are simply stating that in order to obtain an acceptable weak-field limit 
for small $R$, we must choose the constants $a$ and $b$
such that $aR-b  \gg 1$ is satisfied.

\section{Cosmological Evolution}

Having considered the Newtonian limit of our models, we now evaluate
the Friedmann equation, and attempt to reconstruct the cosmic
history of our models. We will find that this model mimics general
relativity very closely throughout its cosmic history.

To show this, we first write down the equivalent Friedmann equation
for a general $F(R)$ model, obtained from the $(0,0)$ component of the field 
equations ($\ref{eq:b2}$),

\begin{equation}\label{eq:b7} -3{\ddot{a} \over a}F'(R) + {F \over 2} + 3H
\partial_{t}\left[F'(R) \right] = {\rho \over M^{2}} .\end{equation}

\noindent With our highly non-trivial $F(R)$, this equation
corresponds to a complex, non linear, third-order differential
equation in the scale factor $a$. However, we can use our
approximations ($\ref{eq:fb3}$) to write ($\ref{eq:b7}$) in a much
simplified form. We first note that the approximations
($\ref{eq:fb3}$) are valid for the entire cosmic history of this
model, because they are applicable for large $\alpha = aR - b$.
Since $R > R_{0}$ throughout the cosmic history, and since $\alpha$
is at its smallest value at $R = R_{0}$, then given that we have
constrained $\alpha$ to be large at $R = R_{0}$ then it must be
large for all $R > R_{0}$.

Using the approximations ($\ref{eq:fb3}$) in ($\ref{eq:b7}$), we
find the following modified Friedmann equation,

\begin{equation}\label{eq:ab7} 3H^{2} - {b \over 2a} +
e^{-2\alpha}\left(3{\ddot{a}\over a} + {1
\over 4a} + 6aH \partial_{t}R \right) + O\left(e^{-4\alpha}\right) =
{\rho \over M^{2}},
\end{equation}

\noindent where we have used the condition that $b$ must be large, and hence $\log(1+\exp(2b)) \approx 2b$. The first two terms are what we would expect from standard
General Relativity with a fine tuned cosmological constant $\Lambda
= {b \over 2a} \approx R_{0}$. The remaining terms on the left hand side are
corrections to the Friedmann equation, but they are highly
suppressed by a factor of $e^{-2\alpha} \ll 1$ for all $ R \geq
R_{0}$. It follows that our model will not significantly differ from
standard GR. The exponentially suppressed corrections to 
general
relativity that we have found are small in the current epoch, and
decrease in significance as we track our model back through its
cosmic history. Additionally, we note that we still have a fine
tuning problem, in that we have two free parameters, and we have
chosen them specifically such that the combination $\Lambda = {b
\over 2a}$ satisfies $\Lambda \ll 1$. We stress however that the
combination $b / 2a$ in ($\ref{eq:ab7}$) is not a true cosmological
constant; rather our model mimics standard general relativity with a
small vacuum energy.

The fine-tuning found in our model is a generic feature of modified
gravity theories. Since we demand that $F''(R) \sim 0$ for small $R$
to obtain an acceptable weak field limit, it follows that in the vicinity of 
the vacuum $R= R_{0}$, we
must have $F''(R) \approx 0$, and hence $F(R) \approx cR + d$, where
$c$ and $d$ are constants. This means that for
vacuum $R \sim R_{0}$, we can expand $F(R)$ in the form $F(R) \approx
cR + d$, where $c$ and $d$ will be some combination of constants in
the model. This is exactly what we have done for our $F(R)$ model
above; we have expanded $F(R)$ as $F(R) \approx R - {b \over a}$
for $R \sim R_{0}$. It follows that to obtain an acceptable vacuum
solution $R_{0} = 12 H_{0}^{2} \ll 1$, we must fine-tune the
parameters in the model.

The fact that our model must necessarily mimic GR
with suppressed corrections is a consequence of the constraints that
we have imposed. Specifically, we have used the fact that to obtain
an acceptable weak field expansion, we require $F''(R) \sim 0$ for
small $R$. In addition, we have also assumed that in the large $R$
regime $F(R)/R \to 1$, which suggests that $F''(R) \sim 0$ as $R \to
\infty$. Since we are considering models which have $F''(R) \neq 0$
for all $R$, then it follows that $F''(R)$ must be small throughout
the cosmic history. 

We expect that any $F(R)$ function that satisfies the consistency conditions that we have considered will possess a Friedmann equation 
of the form ($\ref{eq:ab7}$), that is standard GR plus $\Lambda$ with small correction terms. These correction terms will not necessarily be
exponentially suppressed as they are for the $F(R)$ considered in this paper, but they must be suppressed throughout the expansion history of the Universe regardless of the particular form of $F(R)$.
However, although these models will mimic GR, this does not imply that they are observationally identical to GR if we consider all aspects of gravity. For example, it might be possible to observe
small corrections to GR by considering solar system tests of gravity or the binary pulsar. In
a subsequent paper we hope to find distinct observational signatures
of our model by analyzing local gravity constraints.

Finally, we observe that the conclusions that we have reached do not apply if we were to relax any of the constraints on $F(R)$, for example
by allowing $F'' = 0$ for some $R$ in the past. Hence it might be possible to construct an acceptable model that does not obey all of the constraints at all times,
and hence will not mimic GR throughout the entire cosmic history. In addition, it is not strictly necessary for an acceptable modified gravity model
to obey the consistency conditions for all $R$, since they can only be probed for $R \gtrsim R_{0}$. We could have an acceptable model
which has $F''(R) =0$ for some $R < R_{0}$ in the future. However, there is a possibility of such models evolving to a final state containing instabilities.

\section{\label{sec:5} Conclusion}

We have taken the consistency constraints that exist in the
literature for modified gravity models, and used them to construct a
particular modified gravity function $F(R)$. A study of the model
has shown that in order to satisfy all constraints, it must
act like GR with highly suppressed corrections
throughout cosmic history. In addition, we find that even though
we do not have a true cosmological constant in our setup, we must
still fine-tune the parameters in our $F(R)$ function to obtain an
acceptable late time accelerating epoch. Although we have come to this 
conclusion for our specific model, we expect our
results to hold true for a large class of $F(R)$ models which satisfy all of
the constraints that we have considered.

While we were completing this work we became aware of Refs. \cite{at1,hu2}, 
which come to similar conclusions. In Ref. \cite{at1}, it was found that in 
order to satisfy local gravity constraints, the function $m(r)$ as defined 
in section $\ref{sec:2}$ must satisfy $m(r) \ll 10^{-58}$ for $R \sim 
H_{0}^{2}$. This places a severe constraint upon $F''(R)$, which must be 
very small in the current epoch of the Universe. The model considered in 
this paper can satisfy the constraint $m(r) \ll 10^{-58}$ by making the free 
parameter $b$ sufficiently large. However, such a stringent constraint is 
likely to yield a model that is virtually indistinguishable to GR.

In Ref. \cite{hu2}, a modified gravity function of the form

\begin{equation}\label{eq:c1} \tilde{F}(R) = R - m^{2} {c_{1} \left({R \over 
m^{2}}\right)^{n} \over c_{2} \left({R \over m^{2}}\right)^{n} + 1 } 
\end{equation}

\noindent was considered, where $m$, $n$, $c_{1}$ and $c_{2}$ are free 
parameters. The model considered in this paper shares many similarities with 
($\ref{eq:c1}$); the condition $\tilde{F}'(R) \approx 1$ was imposed in 
Ref. \cite{hu2} both at $R \sim H^{2}_{0}$ and as $R \to \infty$. This 
implies that the function $\tilde{F}(R)$ can be expanded as

\begin{equation}\label{eq:c2} \tilde{F}(R) \approx R - {c_{1} \over 
c_{2}}m^{2} + {c_{1} \over c_{2}^{2}}m^{2} \left({m^{2} \over R}\right)^{n} 
\end{equation}

\noindent for small ${m^{2} \over R} \ll 1 $, which is valid throughout the 
cosmic history of the model. We have arrived at a very similar expansion for 
the $F(R)$ function considered in this paper, ($\ref{eq:fb3}$). The 
difference between the two models is that ($\ref{eq:c2}$) contains power law 
corrections to GR, whereas ($\ref{eq:fb3}$) contains exponentially 
suppressed corrections.


\end{document}